\documentstyle[12pt,epsfig]{article}

\newcommand{\beq}{ \normalsize \begin{eqnarray}}

\newcommand{\eeq}{\end{eqnarray}}

\newcommand{\bea}{ \normalsize \begin{eqnarray}}

\newcommand{\eea}{\end{eqnarray}}

\newcommand{\bd}{\begin{displaymath}}

\newcommand{\ed}{\end{displaymath}}

\newcommand{\gm}{\gamma}

\newcommand{\gev}{\ {\rm GeV}}

\newcommand{\mev}{\ {\rm MeV}}

\newcommand{\prd}{Phys. Rev. D }

\newcommand{\prl}{Phys. Rev. Lett. }

\newcommand{\plb}{Phys. Lett. B }

\newcommand{\jhep}{J. High Energy Phys. }

\newcommand{\npb}{Nucl. Phys. B }

\newcommand{\zpc}{Z. Phys. C }

\begin{document}
\begin{center}
\baselineskip=18pt
 {\huge \bf  The heavy-to-light transitions  }
\vspace{0.4cm} {\huge \bf in the light-cone QCD sum rules
\footnote{Presented by T. Huang, email address: huangt@hptc5.ihep.ac.cn}}\\
\end{center}
\vspace{0.4cm}


\begin{center}Tao Huang$^{a}$, Zuo-Hong
Li$^{b}$ and Xiang-Yao Wu$^{a}$
\end{center}
\begin{center} {\footnotesize a. Institute of High Energy Physics, P.O. Box 918(4),
Beijing 100039, China\footnote{Mailing address}} \\

{\footnotesize b. Department of Physics, Yantai University, Yantai 264005, China}
\end{center}
\renewcommand{\thesection}{Sec. \Roman{section}}
\renewcommand{\thesubsection}{ \arabic{subsection}}
{\vskip 5mm


\begin{center} {\bf Abstract}\end{center} \indent We have analyzed the $B\rightarrow \pi $ and
$B_{s}\rightarrow K$ semileptonic form factors and $B\rightarrow V\gamma
(V=K^*,\rho,\omega)$ processes in the light-cone QCD sum rules. In order to enhance
the predictivity and reliability of numerical results the chiral-current correlator is
employed and the twist-3 light-cone wavefunction can be effectively eliminated from
the sum rules. \vspace{2cm} 


\newpage

\begin{center}
{\bf{\large \bf 1. Introduction}}
\end{center}

$B$ decays into a light meson are an important ground to understand and test the
standard model(SM), since they can provide the signal of CP-violation phenomena and
perhaps, a window into new physic beyond the SM. Nevertheless, the definite
interpretation for the relevant experimental data demand that we have the ability to
precisely compute the physical amplitudes. One is forced to use some approximate
methods, such as QCD factorization formula \cite{1},heavy quark effective
theory(HQET), chiral perturbative theory(CPT), Modified PQCD approach\cite{2}, QCD sum
rules\cite{3} and light cone QCD sum rules\cite{4,5}. Each of them has advantages and
disadvantages. QCD factorization formula is viewed as a great progress in
phenomenology of heavy flavors, however the underlying long distance effects included
in a series of hadronic matrix elements confront still us, which are not being dealt
rigorously with and thus would hinder us from doing such a desired calculation. CHPT
and HQET, as two effective theories at low energy, can describe very well
light-to-light and heavy-to-heavy transitions, respectively. They are not suitable for
a study on heavy-to-light processes. It seems that the modified PQCD approach with a
resummation of Sudakov logarithms is consistent with the physical picture of
heavy-to-light transitions due to the hard gluon exchange. However, a detailed
analysis shows that the reliable PQCD calculation depends on whether the singularities
can be eliminated or suppressed. The singularities come from on shell gluon, on-shell
light quark and on-shell heavy quark. QCD sum rule method is built on the basis of QCD
theory, but the obtained form factors behave very badly in the heavy quark limit
$M_Q\rightarrow \infty$ due to the fact that one omits the effects of the finite
correlation length between the quarks and the gluons in the physical vacuum. In order
to overcome the defect, QCD light cone sum rule(LCSR) approach has been developed in
Ref.\cite{4,5} and has been regarded as an advanced tool to deal with heavy-to-light
exclusive processes. In comparison with the case of the traditional QCD sum rules, in
LCSR approach the operator product expansion(OPE) is carried out near the light cone
$x^2 \approx 0$, instead of in small distance $x \approx 0$, and the non-perturbative
dynamics is parametrized in term of so-called the light cone wavefunctions of light
mesons, instead of the vacuum condensates. It is the striking advantage of the
approach to describe heavy-to-light transitions in a way consistent with the
universally accepted viewpoint that non-perturbative QCD dynamics occupies an dominant
place and perturbative hard gluon exchanges contribute only a subleading effect in
that case. Very recently, it was generalized to study the non-factorizable effects in
$B\to \pi\pi$\cite{6} and to probe heavy-to-light form factors in the whole
kinematically accessible ranges\cite{7}. The technical details of LCSR can be found,
for instance , in \cite{8}, while for a detailed comparison with the traditional sum
rules, see \cite{9}. However, a problem with the approach is that the relevant light
cone wavefunctions are intrinsically non-perturbative and can only be determined via
some phenomenological methods. This could considerably affect the reliability of sum
rule predictions. Accordingly, it is essentially important to find out some certain
way to control the pollution by the light cone wavefunctions. With this end in view, a
pragmatic strategy has been suggested in Ref.[5,10-13], in which some specific chiral
current operators act as the interpolating fields in the correlators used as a LCSR
calculation on the form factors for heavy-to-light transitions, making some twist-3
components cancel out in the OPE. Taking it into account that these twist-3
distributions are equally important in most cases but poorly known to us compared with
the corresponding twist-2 ones, we may think of this procedure as an effective way to
narrow down the uncertainties in LCSR calculations. Here we intend to present a simple
summary of the our previous works [10-13] on applications of the improved LCSR
approach to some important heavy-to-light transition processes.
\begin{center}
{\bf{\large \bf 2. Heavy-to-light Semileptonic Form Factors\cite{10,11}}}
\end{center}

Semileptonic $B$ decays included by the $b\to u$ transition are regarded as the most
promising processes adequate to extract $|V_{ub}|$ from the relevant data. We will
focus on a discussion on $B\to \pi$ and $B\to K$ semileptonic transition form factors
within the framework of an improved LCSR.

The form factors for $B\to \pi$ transitions can be defined as follows,
\begin{eqnarray}
\langle\pi (p)|\overline{u}\gamma _\mu b|B(p+q)\rangle=2f^{B\to \pi}(q^2)p_\mu
+\widetilde{f}^{B\to \pi} (q^2)q_\mu ,
\end{eqnarray}   \label{1}
with $q$ being the momentum transfer. We choose to use a chiral current correlator
\begin{eqnarray} \Pi_\mu (p,q) && =i\int d^4xe^{iqx}\langle\pi (p)|T \{
\overline{u}(x)\gamma _\mu (1+\gamma _5)b(x),\overline{b}(0)i(1+\gamma
_5)d(0)\}|0\rangle \nonumber \\ && =\Pi (q^2,(p+q)^2)p_\mu +\widetilde{\Pi
}(q^2,(p+q)^2)q_\mu ,
\end{eqnarray}            \label{2}
to calculate $f^{B\to \pi}(q^2)$ and $\tilde f^{B\to \pi}(q^2)$. Its hadronic
expansion reads
\begin{eqnarray}
\Pi_\mu ^H(p,q)  &&= \Pi ^H(q^2,(p+q)^2)p_\mu + \widetilde{\Pi}^H(q^2,(p+q)^2)q_\mu
 \nonumber \\
&&= \frac{ \langle \pi|\overline{u}\gamma _\mu b|B\rangle \langle
B|\overline{b}\gamma_5d|0\rangle} {m_B^2-(p+q)^2} \nonumber \\
&&+\sum\limits_H\frac{\langle \pi |\overline{u}\gamma _\mu (1+\gamma _5)|B^H\rangle
\langle B^H|\overline{b}i(1+\gamma _5)d|0\rangle}{m_{B^H}^2-(p+q)^2}.
\end{eqnarray}      \label{3}

We argue that such a procedure is conceptually reasonable and technically feasible by
observing the hadronic form of (2). Because of the special chiral constructs , the
correlator would receive the contribution of the $0^+ B$ mesons, in addition to the
$0^- B$ meson, in its hadronic expression. However, this causes no difficulty in
extracting the desired sum rule for the form factors, Since the lowest $0^+ B$ meson
is far from the ground state of $0^- B$ mesons and slightly below the first excited
$0^- B$ mesons in mass so that we can safely isolate the pole contribution of the
lowest $0^- B$ meson and parametrize them from the higher $0^-$ states as well as all
the $0^+ B$ mesons in a dispersion integral:
\begin{eqnarray}
\Pi^H(q^2,(p+q)^2)=\frac{2 f(q^2) m_B^2 f_B}{m_b (m_B^2-(p+q)^2)} +\int
\limits_{s_0}^{\infty}\frac {\rho^H(s)}{s-(p+q)^2}ds +subtractions,
\end{eqnarray}               \label {4}
\begin{eqnarray}
\widetilde{\Pi} ^H(q^2,(p+q)^2)=\frac{\widetilde{f}(q^2) m_B^2 f_B}{m_b (m_B^2
-(p+q)^2)} +\int\limits_ {s_0}^{\infty}\frac {\widetilde{\rho} ^H(s)}{s-(p+q)^2}ds
+subtractions.
\end{eqnarray}               \label {5}
Here the threshold parameter $s_0$ should be set near the squared mass of the lowest
$0^+ B$ meson.

QCD calculation of the underlying correlator may be allowed, on the other side, for
the negative and large $p^2$ and $(p+q)^2$, in which case the OPE goes effectively in
powers of the deviation from the light cone $x^2\approx 0$. The chiral limit
$p^2=m_{\pi}^2=0$ will be taken throughout this discussion, for simplicity. Carrying
out the OPE for the correlator, we have
\begin{eqnarray}
\Pi^{QCD}(q^2,(p+q)^2)&&=\Pi ^{(\bar{q}q)}[q^2,(p+q)^2]+\Pi
^{(\bar{q}qg)}[q^2,(p+q)^2] \nonumber\\ &&=2f_\pi m_b \left[\int\limits_{0}^{1}
\frac{du}{u}\varphi _{\pi}(u)\frac {1}{s-(p+q)^2}-8m_b^2\int\limits_{0}^{1}\frac{du}{
u^3}g_1(u)\frac{1}{(s-(p+q)^2)^3} \nonumber \right. \\ && \left.
+2\int\limits_{0}^{1}\frac{du}{u^2}
G_2(u)\frac{1}{(s-(p+q)^2)^2}+4\int\limits_{0}^{1}\frac{du}{u^3}G_2(u)\frac{
q^2+m_b^2}{(s-(p+q)^2)^3} \right]\nonumber\\ && + i g_s m_b \int \frac{d^4kd^4xdv}{(2
\pi)^4 (m_b^2-k^2)} e^{i (q-k) x} (\langle\pi(p)|\bar{d}(x)\gamma_\mu G^{\alpha
\beta}(v x) \sigma_{\alpha \beta}u(0)|0\rangle \nonumber \\ && + \langle\pi(p)|\bar{d}
(x)\gamma_\mu\gamma_5G^{\alpha \beta}(v x) \sigma_{\alpha \beta}u(0)|0\rangle ).
\end{eqnarray}               \label {6}
with $\Pi ^{(\bar{q}q)}$ being the two-particle contribution and $\Pi ^{(\bar{q}qg)}$
the three-particle one. $\varphi_\pi (u)$ is the leading twist-2 wavefunction and the
others have twist-4. They are defined as follows
\begin{eqnarray}
\langle\pi (p)|T\bar{u}(x)\gamma _\mu \gamma _5d(0)|0\rangle=&&-ip_\mu f_\pi
\int\limits_0^1due^{iupx}(\varphi _\pi (u)+x^2g_1(u)) \nonumber \\ &&+f_\pi (x_\mu
-\frac{x^2p_\mu }{px})\int\limits_0^1due^{iupx}g_2(u),
\end{eqnarray}\label {7}
\begin{eqnarray}
\langle\pi(p)| \bar{d}(x)\gamma_\mu\gamma_5g_sG_{\alpha \beta}(v x)u(0)|0\rangle
=&&f_\pi\left[q_\beta(g_{\alpha \mu}-\frac{x_\alpha q_\mu}{q x})-q_\alpha (g_{\beta
\mu}-\frac{x_\beta q_\mu}{q x})\right]
 \int D\alpha_i\varphi_\perp(\alpha _i) e^{i q x (\alpha_1+v
\alpha_3)} \nonumber \\ &&+f_\pi \frac{q_\mu}{q x} (q_\alpha x_\beta-q_\beta x_\alpha)
\int D\alpha_i
 \varphi_\parallel(\alpha _i) e^{ i q x(\alpha_1+v \alpha_3)}
 \end{eqnarray}\label {8}
 and
 \begin{eqnarray}
\langle\pi(p)| \bar{d}(x)\gamma_\mu g_s\widetilde{G}_{\alpha \beta}(v x)u(0)|0\rangle
\nonumber= && i f_\pi[q_\beta(g_{\alpha \mu}-\frac{x_\alpha q_\mu}{q x})-q_\alpha
(g_{\beta \mu}-\frac{x_\beta q_\mu}{q x})] \int
D\alpha_i\widetilde\varphi_\perp(\alpha_i) e ^{i q x (\alpha_1+v\alpha_3)} \nonumber
\\
 &&+if_\pi \frac{q_\mu}{q x} (q_\alpha x_\beta-q_\beta
x_\alpha) \int D\alpha_i\widetilde\varphi_\parallel(\alpha_i) e^{i q x (\alpha_1+v
\alpha_3)}.
\end{eqnarray}              \label {9}

At this point,the important observation is that difference from the standard LCSR
calculations, in the present case the twist-3 components cancel precisely out in the
OPE. It can improve greatly the precision of LCSR predictions.

Now the LCSR for $f(q^2)$ can be obtained using the standard procedure. The result is
\begin{eqnarray}
f(q^2)=&&\frac{m_b^2 f_\pi}{m_B^2 f_B} e^{\frac{m_B^2}{M^2}} \left\{\int
\limits_{\triangle}^{1}\frac{du}{u} e^{-\frac{m_b^2-q^2 (1-u)}{u M^2}} \left(\varphi_
\pi(u)-\frac{4 m_b^2}{u^2 M^4} g_1(u) + \frac{2}{u M^2} \int \limits_{0}^{u}g_2(v) dv
(1+\frac{m_b^2+q^2}{u M^2})\right) \nonumber \right. \\
 +&&\int
\limits_{0}^{1} dv \int D\alpha_i\frac {\theta(\alpha_1+v
\alpha_3-\Delta)}{(\alpha_1+v \alpha_3)^2 M^2} e^{-\frac {m_b^2-(1-\alpha_1- v
\alpha_3) q^2}{M^2 (\alpha_1+v \alpha_3)}} (2 \varphi_\perp(\alpha_i)+2
\widetilde\varphi_i\perp(\alpha_i)
-\varphi_\parallel(\alpha_i)-\widetilde\varphi_\parallel(\alpha_i))
\nonumber \\
- &&4 m_b^2 e^{ \frac{-s_0}{M^2}} \left(\frac{1}{(m_b^2-q^2)^2} (1+\frac{s_0-q^2}
{M^2}) g_1(\Delta)-\frac{1}{(s_0-q^2)
(m_b^2-q^2)} \frac {dg_1(\Delta)} {du} \right) \nonumber \\
 -&&\left. 2e^{\frac{-s_0}{M^2}} \left(\frac{m_b^2+q^2}{(s_0-q^2) (m_b^2-q^2)}
g_2(\Delta) - \frac{1}{(m_b^2-q^2)} (1+\frac{m_b^2+q^2}{m_b^2-q^2} (1+\frac{s_0-q^2}
{M^2}) \int \limits_{0}^{\Delta}g_2(v) dv \right) \right\}.
\end{eqnarray}                   \label {10}

To further proceed, we need to make a choice of input parameters.The parameters
entering the sum rule are the $b$ quark mass $m_b$, $B$ meson mass $m_B$, decay
constant $f_B$ and threshold parameter $s_0$. We take $m_b=4.7-4.9\ \gev$, $m_B=5.279\
\gev $. As for the decay constant $f_B$ and the threshold parameter $s_0$, the
two-point correlator
\begin{eqnarray}
K(q^2)=i\int d^4xe^{iqx}\langle0|\overline{q}(x)(1+\gamma_5)b(x),\overline{b}(0)
(1-\gamma_5)q(0)|0\rangle\nonumber
\end{eqnarray}
will be used to estimate it, for consistency. A standard manipulation yields three
self-consistent sets of SR results: (1) $f_B=165\ \mev$ for $m_b=4.7\ \gev$ and
$s_0=33\ \gev^2$, (2) $f_B=120\ \mev$ for $m_b=4.8\ \gev$ and $s_0=32\ \gev^2$, and
(3) $f_B=85\ \mev$ for $m_b=4.9\ \gev$ and $s_0=30\ \gev^2$. The parameters relevant
to the $\pi$ meson contain the decay constant $f_{\pi}$ and the set of light cone
wavefunctions. We use $f_{\pi}=0.132\ \gev$ and the wavefunction models\cite{14,15}
based on the conformal invariance of QCD:
\begin{equation}
\varphi_\pi(u,\mu)=6 u (1-u) [1+a_2(\mu) C_2^{\frac{3}{2}}(2 u-1)+ a_4(\mu)
C_4^{\frac{3}{2}} (2 u-1)+\cdots],\nonumber
\end{equation}
with the coefficients $a_2(u.u_b)=0.35$ and $a_4(u,u_b)=0.18$, at the scale
$u_b=(m_B^2-m_b^2)^{1/2}\approx2.5\gev$, and the Gegenbar polynomials $C_2^{{3}/{2}}
(2 u-1)=\frac{3}{2}\left[5(2u-1) ^2\right]$ and $C_4^{{3}/{2}} (2
u-1)=\frac{15}{8}\left[21 (2u-1) ^4 -14(2u-1)^2+1\right]$, and
\begin{eqnarray}
g_1(u,u_b) && = \frac{5}{2} \varepsilon^2 u^2 \overline{u}^2+\frac{1}{2} \varepsilon
\delta^2 [u \overline{u} (2+13 u \overline{u})+10 u^3 \ln u (2-3 u+\frac{6}{5} u^2)
\nonumber \\&& +10 \overline{u}^3\ln{\overline{u}}(2-3\overline{u}+\frac{6}{5}
 \overline{u}^2)] , \nonumber \\ g_2(u,u_b) && =\frac{10}{3}\delta^2
u\overline{u}(u-\overline{u}).\nonumber
\end{eqnarray}
\begin{eqnarray}
&& \varphi_{\perp}(\alpha_i)=30 \delta^2 (\alpha_1-\alpha_2) \alpha_3^2 [\frac{1}{3}+2
\epsilon (1-2 \alpha_3)] , \nonumber \\ && \widetilde{\varphi}_{\perp}(\alpha_i)=30
\delta^2 \alpha_3^2 (1-\alpha_3) [\frac{1}{3}+2 \epsilon (1-2\alpha_3)] , \nonumber \\
&& \varphi_{\parallel}(\alpha_i)=120 \delta^2 \epsilon (\alpha_1-\alpha_2) \alpha_1
\alpha_2 \alpha_3 , \nonumber \\
&& \tilde{\varphi}_{\parallel}(\alpha_i)=-120 \delta^2 \alpha_1 \alpha_2 \alpha_3
[\frac{1}{3}+\epsilon (1-3\alpha_3)],\nonumber
\end{eqnarray}
with $\delta^2(\mu_b)=0.17\ GeV^2$ and $\varepsilon(\mu_b)=0.36$.

Having fixed the input parameters, we can carry out the numerical analysis. The
reasonable range of $M^2$ is found to be $8\ GeV^2 \leq M^2 \leq 17\ GeV^2$, in which
the variation of $f(q^2)$ with $M^2$ turns out to be negligibly small. For a specific
$M^2=12\ \gev^2$, the sum rules for $f^{B\to\pi}(q^2=0)$ are predicted to $f(0)=0.27,\
0.29$ and $0.33$ (corresponding to the set (3), set (2) and set (1),respectively),
which are in basic agreement with an estimate from the standard LCSR\cite{14}. As a
matter of fact, numerical agreement  between the two different approaches exists up to
$q^2=10\ \gev^2$, the differences being within $20\%$; while for $10\gev^2 \leq q^2
\leq 18\ GeV^2$ there is a numerical disagreement of about $20-25\%$. For the region
$q^2 \geq 18\ GeV^2$, application of the LCSR is questionable such that a comparison
is meaningless between the two sum rule results.

The total uncertainty in $f^{B\to\pi}(q^2)$, which arises from the uncertainties in
the $b$ quark mass $m_b$, the threshold parameter $s_0$, the decay constant $f_B$ and
the pionic light cone wavefunctions, is estimated at the level of about $26\%$.

Following the same procedure, we may analyze the $B_s\to K$ form factor $f^{B_s\to
K}(q^2)$. As compared with the case of $B\to \pi$, however, the $f^{B_s\to K}(q^2)$ is
more difficult to calculate, for SU(3) breaking corrections to the twist-3
wavefunctions of $K$ meson have not been investigated completely in the literature.
Explicitly, this problem can be avoided in our approach.

The correlator used for the LCSR calculation on $f^{B_s\to K}(q^2)$ may be obtained by
an obvious replacement $d\to s$ in Eq. (2). Carrying out the OPE and considering the
non-negligible $K$-meson mass we have
\begin{equation}
\Pi^H(q^2,(p+q)^2)= \frac{2 f^{B_s\to\pi}_{LC}(q^2) m_{B_s}^2 f_{B_s}}{(m_b+m_s)
(m_{B_s}^2-(p+q)^2)} +\int \limits_{s_0}^{\infty}\frac {\rho_1^H(s)}{s-(p+q)^2}ds
\end{equation}
where the definitions $\beta =\alpha_1+\alpha\alpha_3,\
\Delta=(m_b^2-q^2)/(s_0-q^2-m_k^2)$ and $D\alpha_i=d\alpha_1 d\alpha_2d\alpha_3
\delta(1-\alpha_1-\alpha_2-\alpha_3)$ have been used, and the light cone wavefunctions
of the $K$ meson obey the same definitions  as the corresponding cases of $\pi$ meson.

In the following numerical analysis we adopt a model presented in \cite{16},
\begin{equation}
\varphi_K(u)=6u(1-u)\left\{1+1.8\left[(2u-1)^2-1/5\right]-0.5(2u-1)\left[1+1.2[(2u-1)^2-3/7]\right]
\right\}
\end{equation}
for the leading twist-2 wavefunctions of the $K$ meson, and neglect the SU(3) breaking
effects for all the twist-4 distributions. Numerically, we take $f_K=0.16\gev$ and
$m_s=0.15\gev$. Furthermore we use $m_b=4.8\gev$, $m_{B_s}=5.369\gev$,
$m_{B^*}=5.325\gev$, $s_0=34\gev^2$, $f_{B_s}=0.142\gev$, and $f_{B^*}=0.132\gev$ in
the $B$ channel.

With these inputs, the fiducial interval of $M^2$ is determined to be $8\leq M^2\leq
17\gev^2$, depending slightly on $q^2$, for $q^2=0-17\gev^2$. The sum rule results for
$f^{B_s\to K}(q^2)$ are illustrated by the solid line in Fig. 1, the resulting total
uncertainty being about $20\%$.

Having in mind that all the above discussions are limited to the region of low and
intermediate momentum transfer, for which the OPE is effective, one must find another
way to estimate the form factors at the large momentum transfer. The heavy-to-light
transition form factors in the whole kinematical accessible range, for example,
$B_s\to K$ form factor $f^{B_s\to K}(q^2)$ can be precisely be represented as
\begin{eqnarray}
f^{B_s\to K}(q^{2}) &=&\frac{f_{B^{\ast }}g_{B^{\ast }B_{s}K}}{2m_{B^{\ast
}}(1-q^{2}/m_{B^{\ast }}^{2})}+\int\limits_{\sigma _{0}}^{\infty }\frac{\rho (\sigma
)d\sigma }{1-q^{2}/\sigma }  \nonumber \\ &=&F_{G}(q^{2})+F_{H}(q^{2}),
\end{eqnarray}         \label{14}
with $g_{B^{\ast }B_{s}K}$ being the strong coupling defined by $ \langle B^{\ast
}(q,e)K(p)|B_{s}(p+q)\rangle =-g_{B^{\ast }B_{s}K}(p\cdot e)$, and $\rho (\sigma )$ a
spectral function with the threshold $\sigma _{0}$. Obviously, $F_{G}(q^{2})$ stands
for the contribution from the $B^{\ast }$ pole, which describes the principal behavior
of the form factor around $q^{2}=q_{max}^{2}$, and $F_{H}(q^{2})$ parametrizes the
higher state effects in the $B^{\ast }$ channel. The non-perturbative parameter
$f_{B^{\ast }}g_{B^{\ast }B_{s}K}$ is calculable with the correlator used for the LCSR
estimate of $f_{LC}^{B_s\to K}(q^{2})$. Accordingly, modelling the higher state
contributions by a certain assumption and then fitting Eq.(14) to its LCSR result
$f^{B_s\to K}(q^{2})$ in the region accessible to the light cone OPE, we might derive
the form factor $f^{B_s\to K}(q^{2})$ in the total kinematical range with a better
accuracy.

We work in the large space-like momentum regions $(q^{2}\ll 0$ and
$(p+q)^{2}\ll 0)$ for the correlator in question. Making the Borel
improvements on the yielded theoretical expression
$F^{QCD}(q^{2},(p+q)^{2})\rightarrow \bar{F}^
{QCD}(M_{1}^{2},M_{2}^{2})$, and then matching it onto the
corresponding Borel improved hadronic form via the use of
quark-hadronic duality ansatz, the final sum rule for
$f_{B^*}g_{B^{\ast }B_{s}K}$ reads,
\begin{eqnarray}
f_{B^{\ast }}g_{B^{\ast }B_{s}K} &=&\frac{2m_{b}(m_{b}+m_{s})f_{K}}{
m_{B_{s}}^{2}m_{B^{\ast }}}e^{\frac{m_{B_{s}}^{2}+m_{B^{\ast }}^{2}}{2
\overline{M}^{2}}}\left\{ \overline{M}^{2}\left[ e^{-\frac{m_{b}^{2}+\frac{1
}{4}m_{K}^{2}}{\overline{M}^{2}}}-e^{-\frac{S_{0}}{\overline{M}^{2}}}\right]
\varphi _{K}(1/2)\right.  \nonumber \\
&&\left. +e^{-\frac{{m_{b}^{2}+\frac{1}{4}m_{K}^{2}}}{{\overline{M}^{2}}}} \left[
g_{2}(1/2)-\frac{4m_{b}^{2}}{\overline{M}^{2}}\left[
g_{1}(1/2)-\int\limits_{0}^{1/2}g_{2}(v)dv\right] \right. \right.  \nonumber\\
&&\left. \left. +\int\limits_{0}^{1/2}d\alpha _{1}\int\limits_{1/2-\alpha
_{1}}^{1-\alpha _{1}}\frac{d\alpha _{3}}{\alpha _{3}}\left[ 2\varphi _{\perp }(\alpha
_{i})+2\widetilde{\varphi }_{\perp }(\alpha _{i})-\varphi _{\parallel }(\alpha
_{i})-\widetilde{\varphi }_{\parallel }(\alpha _{i}) \right] \right] \right\} .
\end{eqnarray}
The resulting sum rule for $f_{B^{\ast }}g_{B^{\ast }B_{s}K}$ is numerically
$f_{B^{\ast }}g_{B^{\ast }B_{s}K}=3.88\gev$, with the uncertainty of about $18\%$.
Using its central value,we obtain a $B^{\ast}$ pole approximation for $f^{B_s\to
K}(q^{2})$, which is shown by the dotted line in Fig.1. It is explicitly demonstrate
that a perfect match between the direct LCSR and $B^{\ast}$ pole predictions appears
at $q^2\approx 15-20\gev^2$.

It is important and interesting to make a comparison of our sum rule results and those
from the standard LCSR based on the correlator vector and pseudoscalor currents, which
are easy to obtain using the twist-3 wavefunction suggested in Ref. \cite{17}, leaving
the twist-4 distribution amplitudes unchanged and making a corresponding replacement
of the other relevant input parameters in Eq.(79) and (44) of Ref. \cite{14}.We
observe that the standard approach gives the same matching range as in our case and
the resulting deviations from our predictions turn out to be between $-10\%--15\%$,
depending on $q^{2}$, in the total kinematically accessible region.

Assuming the higher state contribution in Eq.(14) to obey
$F_H(q^2)=a/\left(1-bq^2/m_{B^{*}}^2-cq^4/m_{B^{*}}^4\right) $, we could give a model
for the form factor
\begin{eqnarray}
f^{B_s\to K}(q^{2})=\frac{f_{B^{\ast }}g_{B^{\ast }B_{s}K}}{2m_{B^{\ast
}}(1-q^{2}/m_{B^{\ast }}^{2})}+\frac{a}{1-bq^{2}/m_{B^{\ast }}^{2}-cq^{4}/m_{B^{\ast
}}^{4}}.
\end{eqnarray}
The parameter $a$ can easily be fixed at $-0.07$, using the central values of
$f^{B_s\to K}_{LC}(0)$ and $f_{B^{\ast }}g_{B^{\ast }B_{s}K}$. In the region $
q^{2}=0-18\ GeV^{2}$, the best fit of Eq.(16) to $f_{LC}(q^{2})$ yields $b=1.11$ and
$c=-8.33$. It turns out that the fitting results(the dashed line in Fig.1) reproduce
precisely the LCSR prediction up to $q^{2}=18\ GeV^{2}$ and support considerably the
$B^{\ast}$ pole description of the $ B_{s}\rightarrow K$ form factor in large $q^{2}$
region.

Finally, we look roughly into SU(3) breaking effects in heavy-to-light decays by
considering the ratio of the derived $f^{B_s\to K}(q^{2})$ over $f^{B_s\to
\pi}(q^{2})$, which can be modelled using all the same method as in the $B_s\to K$
case. For the common kinematical region the resulting ratios, a comparable result
$1.05-1.15$ with that from the standard approach, favor a small SU(3) breaking effect.
\begin{center}
{\bf{\large \bf 3. The rare decays $B\to (K^*,\rho,\omega)+\gamma$}\cite{12,13}}
\end{center}

The rare decays $B\to (K^*,\rho,\omega)+\gamma$ induced by the flavor-changing neutral
currents (FCNC's) provide us with a good opportunity to search for new physics. These
decay modes are dominated by the electromagnetic penguin operators responsible for
$b\to (s,d)+\gamma$. The relevant form factors $F^{B\to (K^*,\rho,\omega)}(0)$ have
been studied widely by using the various tricks. With the standard LCSR, ones found
$F^{B\to K^*}(0)=0.32+0.05$ \cite{18} and $F^{B\to \rho}(0)=0.285\pm15\%$\cite{19}.

Here we reanalyze them using the improved LCSR approach, to eliminate the pollution by
some of nonleading distributions. First, we focus on the case of $B\to
K^{\ast}\gamma$. The relevant decay amplitude reads
\begin{eqnarray}
A(B \to K^*\gamma)=Cm_b\epsilon^{\mu}<K^*(p,\eta)|\bar s
\sigma_{\mu\nu}(1+\gamma_5)q^{\nu}b|B(p+q)>,
\end{eqnarray}
where $\epsilon$ and $q$ are the emitted photon polarization vector and momentum,
respectively. The constant $C$ depends on the CKM matrix elements $V_{ts}^* V_{tb}$
and its apparent form can be found in Ref.\cite{20}.

The hadronic matrix element in \cite{7} may be parametrized in terms of the form
factor $F^{B\to K^*}(0)$,
\begin{eqnarray}
<K^{*}(p,\eta)|\bar{s}\sigma_{\mu\nu}(1+\gamma_5)q^{\nu}b|B(p+q)>
                 =(-2i\epsilon_{\mu\nu\alpha\beta}
                 \eta^{*\nu}q^{\alpha}p^{\beta}
                 +2p\cdot{q}\eta^{*}_{\mu}
                 -2q\cdot{\eta^{*}p_{\mu}})F^{B\to K^{\ast}}(0).
\end{eqnarray}
A correlator, which is suitable for our purpose, is
\begin{eqnarray}
F_{\mu}(p,q) &&=i\int d^{4}xe^{iqx}<K^{*}(p,\eta)|T\bar{s}(x) \sigma_{\mu\nu}
(1+\gamma_5) q^{\nu}b(x),\bar{b}(0) i(1+\gamma_5)u(0)|0>\nonumber\\
&&=-2\left[i\epsilon_{ \mu\nu\alpha\beta}\eta^{\ast\nu}q^{\alpha}p^{\beta}-p\cdot
q\eta_{\mu}^{\ast}+q\cdot \eta^{\ast} p_{\mu}\right]F\left[(p+q)^2\right].
\end{eqnarray}
The hadronic expression for the invariant function $F\left[(p+q)^2\right]$ is of the
following form
\begin{equation}
F^H\left[(p+q)^2\right]=\frac{m_Bf_B F^{B\to K^{\ast}}(0)}{m_b\left[m_B^2-(p+q)^2
\right]}+\int \limits_{s_0}^{\infty}\frac {\rho^H(s)}{s-(p+q)^2}ds.
\end{equation}
Here the dispersion integral stands for the contribution of both higher pseudoscalar
states $B_p^h$ and scalar resonance states $B_s^h$.

On the other hand, for the region of large spacelike momenta $(p+q)^2\ll0$ the
resulting QCD form of $F\left[(p+q)^2\right]$, to the twist-2 accuracy, reads,
\begin{equation}
F^{QCD}\left[(p+q)^2\right]=m_b\int \limits_{m_b^2}^{\infty}\frac {\varphi_{\perp}
(u,u_b^2)}{s-(p+q)^2}\frac{u}{u^2m_{K^{\ast}}^2+m_b^2}ds.
\end{equation}
The variable $u$, the fraction of the $K^{\ast}$ meson momentum carried by the $s$
quark, is related to $s$ by $s=m_b^2/u-(1-u)m_{K^{\ast}}^2$. The leading twist-2
wavefunction $\varphi_{\perp} (u,u_b^2)$ parametrizes the nonlocal matrix element
$<K^{*}(p,\eta)|\bar{s}(x)\sigma_{\mu\nu}q^{\nu}(1+\gamma_5)u(0)|0)>$ as the
following,
\begin{eqnarray}
<K^{*}(p,\eta)|\bar{s}(x)\sigma_{\mu\nu}q^{\nu}(1+\gamma_5)u(0)|0>
&&=i\left[(q\cdot\eta^{*})p_{\mu}-p\cdot{q}\eta^{*}_{\mu}-\epsilon_{
\mu\nu\alpha\beta}e^{\nu}q^{\alpha}p^{\beta}\right]f^{K^{\ast}}_{\perp}\nonumber\\
&&\times\int\limits_{0}^{1}due^{iup\cdot x}\varphi_{\perp}(u,u_b^2),
\end{eqnarray}
with the decay constant $f^{K^{\ast}}_{\perp}=210\mev$.

The final LCSR for the form factor $f^{K^{\ast}}(0)$ is
\begin{eqnarray}
\frac{m_B^2}{m_b}f_{B}F^{B\to K^*}(0)e^{\frac{-(m_{B}^{2}-m_{b}^2)}{T}}
=\int_{u(s_0)}^{1}\frac{m_{b}f_{\perp}}{u}{\varphi_{\perp} (u,\mu^2)}e^{\frac{u-1}{T}
(\frac{m_b^2}{u}+m_{K^*}^2)},
\end{eqnarray}
which, in comparison with that in Ref.\cite{18}, does not receive the contribution of
distribution amplitudes $\varphi_{\parallel}^0,\ g^v_{\perp}$ and $g^a_{\perp}$, and
thus is more reliable.

Using as inputs the Chernyak-Zhitnitsky(CZ) model for
$\varphi^{K^{\ast}}_{\perp}(u,\mu^2)$,
\begin{eqnarray}
\varphi_{\perp}^{K^{\ast}}(u,\mu^2)=6u(1-u)\bigg[1+0.57\omega-1.35(\omega^2-\frac{1}{5})
  +0.46(\frac{7}{3}\omega^3-\omega)\bigg].
\end{eqnarray}
with $\omega=2u-1$, and taking into account all the possible uncertainties in the
numerical analysis, we have $F^{B\to K^{\ast}}(0)=0.34\pm0.05$, slightly greater than
that in Ref.\cite{18} and $B(B\to K^{\ast}\gamma)=(5.1\pm1.7)\times 10^{-5}$, a result
comparable with the experimental observation\cite{21} $B_r(B\to
K^{\ast}\gamma)=(4.5\pm1.5\pm0.9)\times 10^{-5}$.

The corresponding relations in $B\to (\rho,\omega)+\gamma$ cases can be obtained by
making replacements $s\to d$, $\varphi^{K^{\ast}}_{\perp}(u,\mu^2)\to
\varphi^{(\rho,\omega)}_{\perp}(u,\mu^2)$, $m_{K^{\ast}}\to m_{(\rho,\omega)}$ and
$f_{\perp}^{K^{\ast}}\to f_{\perp}^{(\rho,\omega)}$. The isospin symmetry allows us to
adopt the same wavefunction model for $\rho^+,\ \rho^-,\ \rho^0$ and $\omega$. We
choose the two different models
\begin{eqnarray}
\varphi_{\perp}(u,\mu^2)=6u(1-u)\bigg[1+0.138\times75(\omega^2-0.2)\bigg]
\end{eqnarray}
and
\begin{eqnarray}
\varphi_{\perp}(u,\mu^2)=6u(1-u)\bigg[1+0.077\times7.5(\omega^2-0.2)-0.077\times\left(39.375
\omega^4-26.25\omega^2+1.875\right)\bigg],
\end{eqnarray}
which are respectively introduced by Ball and Braun in Ref.\cite{22} and Bakulev and
Mikhailo in Ref.\cite{23}, and called the model I and the model II. In the two cases,
the resulting sum rule predictions are summarized as follows:
\begin{equation}
F^{B\to\rho^{\pm}}(0)=0.335\pm0.050, \ Br(B\to\rho^{\pm}\gm)=(2.71\pm1.00)
\times10^{-6} \nonumber
\end{equation} and
\begin{equation}
F^{B\to\rho^0/\omega}(0)=0.237\pm0.035, \ Br(B\to(\rho^0, \omega)+\gamma)=
(1.36\pm0.50) \times10^{-6}, \nonumber
\end{equation}
for the model I, and
\begin{equation}
F^{B\to\rho^{\pm}}(0)=0.272\pm0.029,\ Br(B\to\rho^{\pm}\gm)=(1.79\pm0.61)
\times10^{-6} \nonumber\\
\end{equation}
and
\begin{equation}
F^{B\to\rho^0/\omega}(0)=0.192\pm0.021,\ Br(B\to(\rho^0,
\omega)+\gamma)=(0.90\pm0.31)\times10^{-6}, \nonumber
\end{equation}
for the model II. The results with the model I are slightly great than those in Ref.
\cite{19}, while the ones with the model II accord with those in Ref.\cite{19}.
Compared with the experimental observation Ref.\cite{24} $Br(B^0\to\rho^{0}\gamma)
\leq3.9\times 10^{-5}$, $Br(B^0\to\omega\gamma)\leq1.3\times 10^{-5}$ and
$Br(B^-\to\rho^{-}\gamma)\leq1.1\times 10^{-5}$, our predictions are below the
experimental upper limits.
\begin{center}
{\bf{\large \bf 4. Conclusion}}
\end{center}

Light-cone QCD sum rules is a good framework for calculating heavy-to-light form
factors.We have analyzed the. $B \to \pi$ and $B_s \to K$ semileptonic form factors,
also $B \to K^* \gamma$ and $B \to (\rho, \omega) \gamma$ processes. In order to
enhance the predictivity and reliability of numerical results it is worthwhile to
study how to reduce the various uncertainties in the light-cone QCD sum rules. Here we
employ the chiral-current correlator. It is explicitly shown that the twist-3
light-cone wavefunction, which have not been understood well, can be effectively
eliminated from the sum rules.

\newpage
\begin{figure}
\centerline{ \epsfxsize=12cm \epsfbox{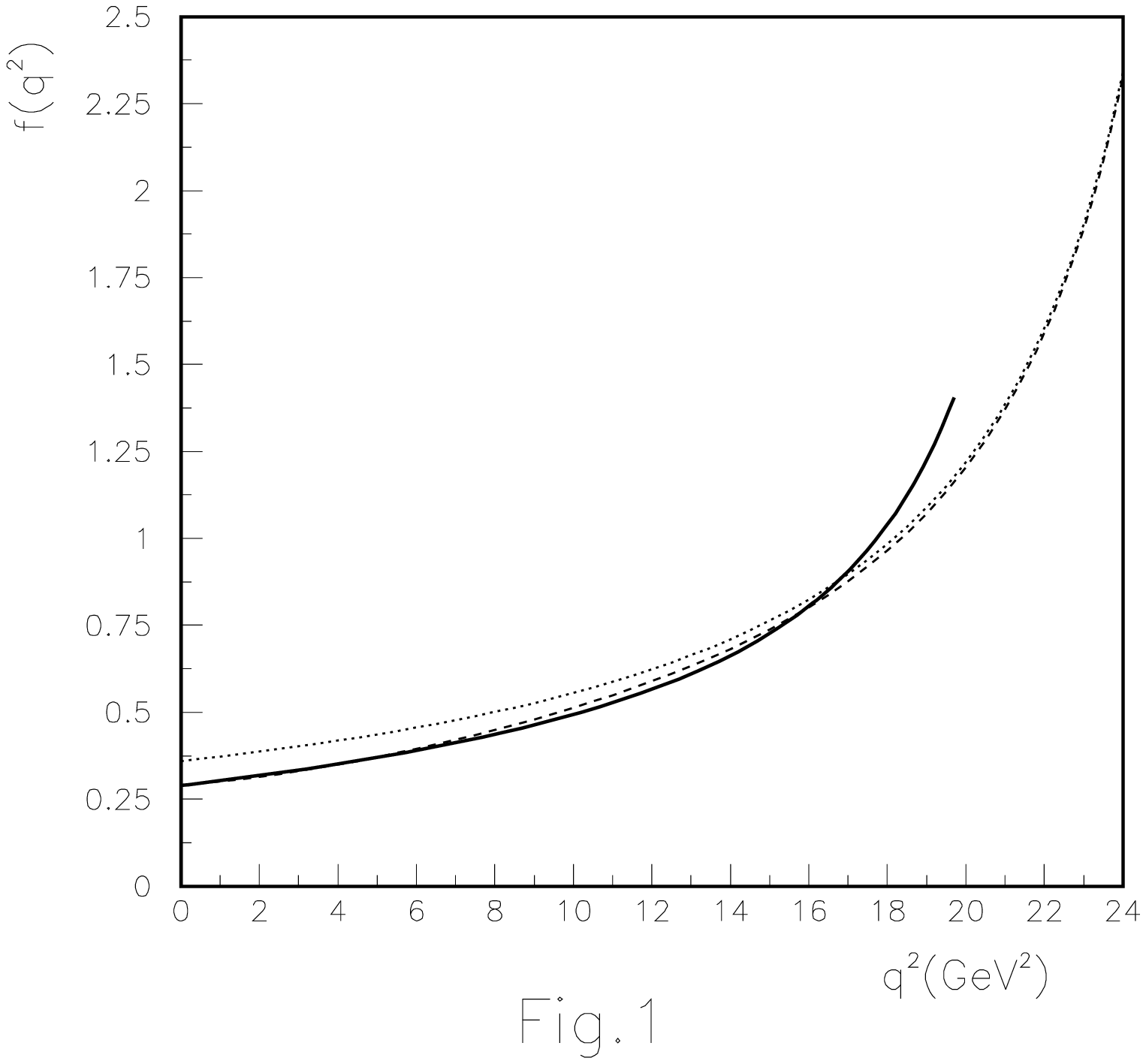} } \caption{\baselineskip 20pt The
$B_s\rightarrow K$ form factor $f(q^2)$ in the total kinematical range. The solid line
denotes the LCSR result $f_{LC}(q^2)$, which is reliable for $0\leq q^2 \leq 17\
GeV^2$. The dotted line expresses the $B^*$ pole prediction suitable for large $q^2$.
The best fit of Eq. (11) to $f_{LC}(q^2)$ is illustrated by the dashed line. It should
be understood that the plotted curves correspond to the central values of all the
relevant parameters.}
\end{figure}
\end{document}